\documentclass[twocolumn,nofootinbib]{revtex4-1}
\usepackage{graphicx}
\usepackage{amsmath, amsthm, amssymb}
\usepackage{slashed}
\usepackage{url}
\usepackage[pdftex]{hyperref}
\usepackage{color}
\usepackage{bm}
\usepackage{mathtools}
\usepackage{subfigure}

\newcommand{\phivev}{\langle\phi\rangle}

\newcommand{\bea}{\begin{eqnarray}}
\newcommand{\eea}{\end{eqnarray}}

\begin{document}

\preprint{UCI-HEP-TR-2018-15}

\title{Early Cosmological Period of QCD Confinement}
\author{Seyda Ipek}
\author{Tim M.P. Tait}
\affiliation{Department of Physics and Astronomy, University of
California, Irvine, CA 92697-4575 USA}
\date{\today}
\begin{abstract}
If the strong coupling is promoted to a dynamical field-dependent quantity, it is possible that the strong force looked very different in the early Universe.
We consider a scenario in which the dynamics is such that QCD confines at high temperatures with a large dynamical scale, relaxing back to
$\sim 1$~GeV before big bang nucleosynthesis.  We discuss the cosmological implications and explore
potential applications, including fleshing out a new mechanism for baryogenesis which opens up if QCD confines before the electroweak
phase transition of the Standard Model.
\end{abstract}

\maketitle
\section{Introduction}
\label{sec:intro}

The Standard Model (SM) of particle physics provides a fantastic description of a plethora of low energy observations.
That said, it remains incomplete, and experimental probes to date have been largely limited to low energies and temperatures.
As a result, it is an intriguing possibility that there could be new physics operating at early cosmological times when the Universe
was much hotter than it is today.

The SM predicts that QCD deconfines at temperatures $T\gtrsim$~GeV
and the electroweak symmetry is restored at temperatures $T\gtrsim$~100~GeV.  But our precise understanding of the cosmological
history becomes fuzzy for temperatures $T \gtrsim 10~{\rm MeV}$ when big bang nucleosynthesis (BBN) begins \cite{Tanabashi:2018oca}.  
It is thus entirely possible that
there is physics beyond the SM which produces a radical departure from the standard cosmological picture at earlier times.

Given this blind spot, we ask the question: what if the scale of QCD confinement was itself varying in the early Universe, settling down to the $\sim$~GeV value
we observe today sometime before BBN?  We introduce dynamics which allow us to explore this possibility, and map out the consequences for
cosmology, the electroweak phase transition, and an opportunity to explain the baryon asymmetry of the Universe.

\section{Dynamical QCD Coupling}
\label{sec:dynamicalQCD}

In order to promote the QCD coupling to a dynamical quantity, we introduce a scalar field $\phi$, taken to be a SM singlet, coupled to the gluon
field strength $G^{\mu\nu}$ via,
\begin{align}
 \mathcal{L}\supset -\frac14 \left( \frac{1}{g_{s0}^2}+\frac{\phi}{M_*} \right)G_{\mu\nu}G^{\mu\nu},  ~~
 \label{eq:gsphi}
\end{align}
where $g_{s0}$ is the QCD coupling for $\langle \phi \rangle = 0$,
and $M_*$ encodes the short distance physics mediating the interaction. 
$\phi$ could represent fluctuations in a radion or a dilaton, or could represent a generic scalar field
which couples to gluons via a triangle diagram containing heavy vector-like colored particles. 
In that case $M_* \sim {4\pi M_Q} / {n_Q y_Q \alpha_s}$, where $n_Q$ is the number of colored fermions with mass $M_Q$
and Yukawa coupling $y_Q$. 
It would be interesting to explore the details of the UV dynamics in more detail, but we leave that for future work.

\subsection{Strong Coupling and the QCD Scale}

When $\phi$ acquires a vacuum expectation value (VEV), it renormalizes the wave function of the gluons, modifying the
effective strong coupling.  In addition, the coupling $g_{s0}$ runs with the renormalization scale $\mu$ in the usual way.  At one loop,
\begin{align}
\frac{1}{\alpha_s(\mu,\phivev)}=\frac{33-2 n_f}{12\pi}\ln\left(\frac{\mu^2}{\Lambda_0^2}\right)+4\pi\frac{\phivev}{M_*},
\end{align}
where $n_f$ is the number of quark flavors with masses $m_f \ll \mu$, and the scale 
$\Lambda_0$ encodes the value of $g_{s0}$ at some UV reference scale.

Confinement is triggered at the scale $\Lambda$ for which ${\alpha^{-1}_s(\Lambda,\phivev)}\simeq 0$,
\begin{align}
\Lambda(\phivev) = \Lambda_0 ~{\rm Exp} \left( \frac{24\pi^2}{2n_f-33}\frac{\phivev}{M_*} \right).
\label{eq:newQCDscale}
\end{align}
For $n_f=6$, $\Lambda_0\sim$~GeV and 
$\phivev/M_*=-0.62$, QCD confines at $\Lambda\sim1$~TeV, well above the temperature of the usual electroweak phase transition.

\subsection{$\phi$ Potential}

We consider a generic potential for $\phi$ at zero temperature, 
\begin{align}
V(\phi)=\alpha_1 \phi +\alpha_2 \phi^2 +\alpha_3 \phi^3 +\alpha_4 \phi^4 + \beta_1 \phi h^\dagger h+\beta_2 \phi^2 h^\dagger h,
\end{align}
where $\alpha_i$ and $\beta_i$ are couplings with appropriate mass dimension.  
We have included a trilinear and mixed quartic with the SM
Higgs doublet $h$, which would modify the $\phi$ dynamics, and lead to
mixing between $\phi$ and the SM Higgs, and are bounded by LHC measurements \cite{Khachatryan:2016vau}.  Since such
interactions do not change the qualitative picture, we simplify the analysis by assuming that they are negligibly small.

When QCD confines, the interaction in Eq.~(\ref{eq:gsphi}) further induces a non-perturbative contribution to $V(\phi)$ via the gluon
condensate, $\langle GG\rangle \propto \Lambda^4 (\phivev)$ \cite{vonHarling:2017yew}.  
For potentials with a sizable $\alpha_1$, this effect is typically unimportant.
For example, a benchmark point with $\alpha_1={\rm TeV}^3,~\alpha_2={\rm TeV}^2,~\alpha_3={\rm TeV}$, $\alpha_4=0.1$, and
$M_* \sim 13$~TeV results in $\phivev \sim -7$~TeV and $m_\phi\sim$~4~TeV, with a 
negligible contribution from the gluon condensate.
 
\subsection{Temperature Dependence} 

In order to restore ordinary QCD with $\Lambda \sim 1$~GeV at low temperatures, $\phivev$ must shift by an amount of
order $\sim 1/2 M_*$.  The details of how this occurs depend sensitively on both the
zero temperature $V(\phi)$ and the finite temperature corrections to it.   Whatever the mechanism, it
is clear that successfully realizing the BBN
predictions for the primordial abundances of the light elements
requires that the temperature at which $\Lambda$ reaches $\Lambda_{\rm QCD}$ satisfy $T_{\rm res} \gtrsim 10$~MeV.

There are a number of constructions which could accomplish this temperature dependence.
\textbf{(i)} There may be species (which could be SM singlets) in the plasma with significant coupling to $\phi$, contributing to its effective mass $\propto g^2 T^2$. Such species could play an important role in $\phi$ phenomenology. \textbf{(ii)} If Eq.~(\ref{eq:gsphi}) is generated by vector-like quarks, there may be additional temperature-dependent contributions to their masses, suppressing the interaction at low temperatures. \textbf{(iii)} There could be a scalar field $\psi$ with its own coupling $(\psi / M_*) GG$, but whose potential induces a positive VEV with parameters tuned to partially cancel the $\phi$ contribution. \textbf{(iv)} Additional scalar fields could couple to $\phi$, and themselves undergo symmetry-breaking at low temperatures, triggering a shift in the effective $V(\phi)$ (see, \emph{e.g.}, \cite{Baker:2016xzo}). For example,
one could introduce two real singlet scalar fields $\phi$ and $\psi$ with the zero-temperature potential
\begin{align}
V(\phi,\psi)=~ &\alpha_1 \phi +\alpha_2 \phi^2 +\alpha_3 \phi^3 +\alpha_4 \phi^4 \notag\\
& + \beta_1 \psi^2 +\beta_2\psi^4 + \gamma_1 \phi\psi^2+\gamma_2\phi^2\psi^2, \label{eq:SBpotential}
\end{align}
which is invariant under $Z_2$ symmetry $\psi \to -\psi$. 
For an appropriate choice of parameters, at high temperatures the fields $\phi$ and $\psi$ have zero VEVs. 
As the universe cools down, first $\phi$ acquires a VEV, and then at a lower temperature, $\psi$ acquires a VEV, which triggers 
the transition $\langle \phi \rangle \to 0$. 
 
In the remainder of this work, we remain agnostic concerning the nature of the physics which provides the necessary temperature dependence in $V(\phi)$, 
simply assuming that some mechanism restores standard QCD somewhere in the range GeV~$\lesssim T_{\rm res }\lesssim 100$~GeV.


\section{Phase Transition and Electroweak Symmetry Breaking}
\label{sec:qcdtransition}

\begin{figure}[h]
\includegraphics[width=.45\textwidth]{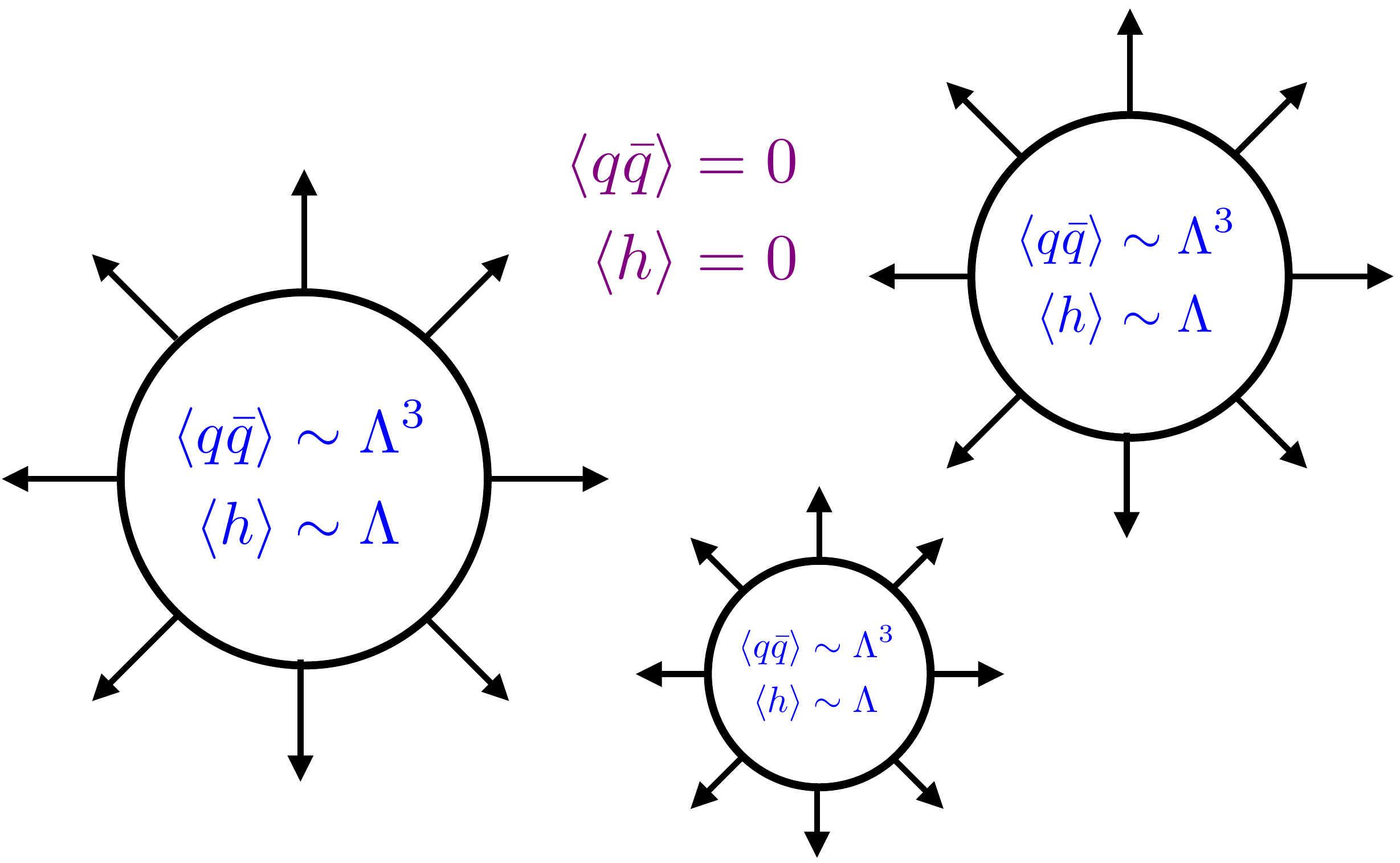}
\caption{Bubbles of confined QCD phase are generated and expand. Inside the bubble, the Higgs 
acquires a VEV due to the tadpole term induced via the quark condensate.} \label{fig:QCDbubbles}
\end{figure}

If $SU(3)$ confines before the electroweak phase transition,
it triggers electroweak symmetry breaking via chiral symmetry breaking.
For $n_f = 6$ massless quark flavors, the chiral phase transition is expected to be strongly first order \cite{PhysRevD.29.338, 1742-6596-432-1-012027},
and proceeds by nucleating bubbles of confined phase with $\langle \bar{q} q \rangle \neq 0$, which expand
to fill the Universe.
The chiral condensate also couples to the Higgs via the quark Yukawa interactions, appearing as a tadpole which 
induces a Higgs VEV inside the bubbles.  This picture is illustrated in cartoon form in Figure~\ref{fig:QCDbubbles}.

The precise details of the QCD phase transition, bubble nucleation, bubble profile, and expansion,
are non-perturbative and beyond the scope of this work \cite{KAJANTIE1992331, Bai:2018dxf}.   
Here, we model the dynamics
by a linear sigma model reflecting the approximate 
$SU(6)_L \times SU(6)_R$ flavor symmetry of QCD,
which is explicitly broken by the SM Yukawa interactions.
The field $\Pi (x)$ is a $6 \times 6$ complex scalar containing the pions, 
scalar mesons, and chiral symmetry-breaking VEVs which is taken
to transform under $SU(6)_L \times SU(6)_R$ as
\bea
\Pi (x) \rightarrow L ~\Pi(x) ~R^\dagger
\eea
where $L$ and $R$ are $SU(6)_{L,R}$ transformations, respectively\footnote{Note that the 
electroweak $SU(2)_W$ is a subgroup of $SU(6)_L$.}.  Below the confinement scale, 
the dynamics of QCD can be described by an effective field theory
containing $\Pi (x)$ and the baryons (which are not important for this discussion),
\begin{align}
{\cal L}_\Pi  = ~ &{\rm Tr} \left[ \partial^\mu \Pi^\dagger \partial_\mu \Pi \right] + \mu^2 ~{\rm Tr} \left[ \Pi^\dagger \Pi \right] 
- \lambda_2 ~{\rm Tr} \left[ \Pi^\dagger \Pi \right]^2  \nonumber \\
&- \lambda_1 ~{\rm Tr} \left[ \Pi^\dagger \Pi \Pi^\dagger \Pi \right]  + \frac{1}{M^2} ~{\rm Det}\, \Pi + H.c.
\end{align}
Naive dimensional analysis (NDA) \cite{Cohen:1997rt} suggests that up to ${\cal O}(1)$ numbers,
$\mu^2 \sim \bar{\Lambda}^2$, $M \sim \bar{\Lambda} / g^2$, and $\lambda_1 \sim \lambda_2 \sim g^2$, where $g \sim 4 \pi$ and $\bar{\Lambda}$ 
is the cut-off of the chiral effective theory (typically of the order of the $\rho$-meson mass.) 
Neglecting the SM Yukawa interactions, this would result in a VEV for $\Pi$, breaking $SU(6)_L \times SU(6)_R \rightarrow SU(6)_D$, of the form
$\langle \Pi^i_j \rangle = f_\pi \delta^i_j$ where $f_\pi^2 \sim \bar{\Lambda}^2 / g^2$.

The explicit $SU(6)_L \times SU(6)_R$ breaking from the quark Yukawa interactions can be included by treating $Y = y h$ as a spurion,
where $h$ is the neutral CP even component of the Higgs doublet, and
in the diagonal quark mass basis $y$ is the $6 \times 6$ diagonal matrix whose entries are the quark Yukawa interactions.
The corresponding terms containing the spurion $Y$ read,
\begin{align}
{\cal L}_Y  =~ & (\tilde{m}^2 \,{\rm Tr} \left[ \Pi^\dagger Y \right] + {\rm h.c.})- \tilde{\lambda}_1 {\rm Tr} \left[ \Pi^\dagger Y \Pi^\dagger Y \right] \notag \\
& - \tilde{\lambda}_2 {\rm Tr} \left[ \Pi^\dagger Y Y^\dagger \Pi \right]
 - \tilde{\lambda}_3 {\rm Tr} \left[ \Pi Y^\dagger Y \Pi^\dagger \right] 
\end{align}
with NDA estimates $\tilde{m}^2 \sim \bar{\Lambda}^2 /g$, $\tilde{\lambda}_1 \sim \tilde{\lambda}_2 \sim \tilde{\lambda}_3 \sim 1$.
The first term expands into a tadpole for $h$ induced by the non-zero $\langle \bar{q} q \rangle$ condensate.  The remaining terms
induce masses\footnote{The pion masses in a given epoch scale with both $\Lambda$ and $\langle h \rangle$.} 
for the 35 pions and
induce a back-reaction where the Higgs VEV reduces the corresponding chiral condensate by producing a mass for the SM quarks. 
In principle, this results in a complicated set of coupled equations, however a simple estimate provides a heuristic picture.

\begin{figure}[t]
\includegraphics[width=.4\textwidth]{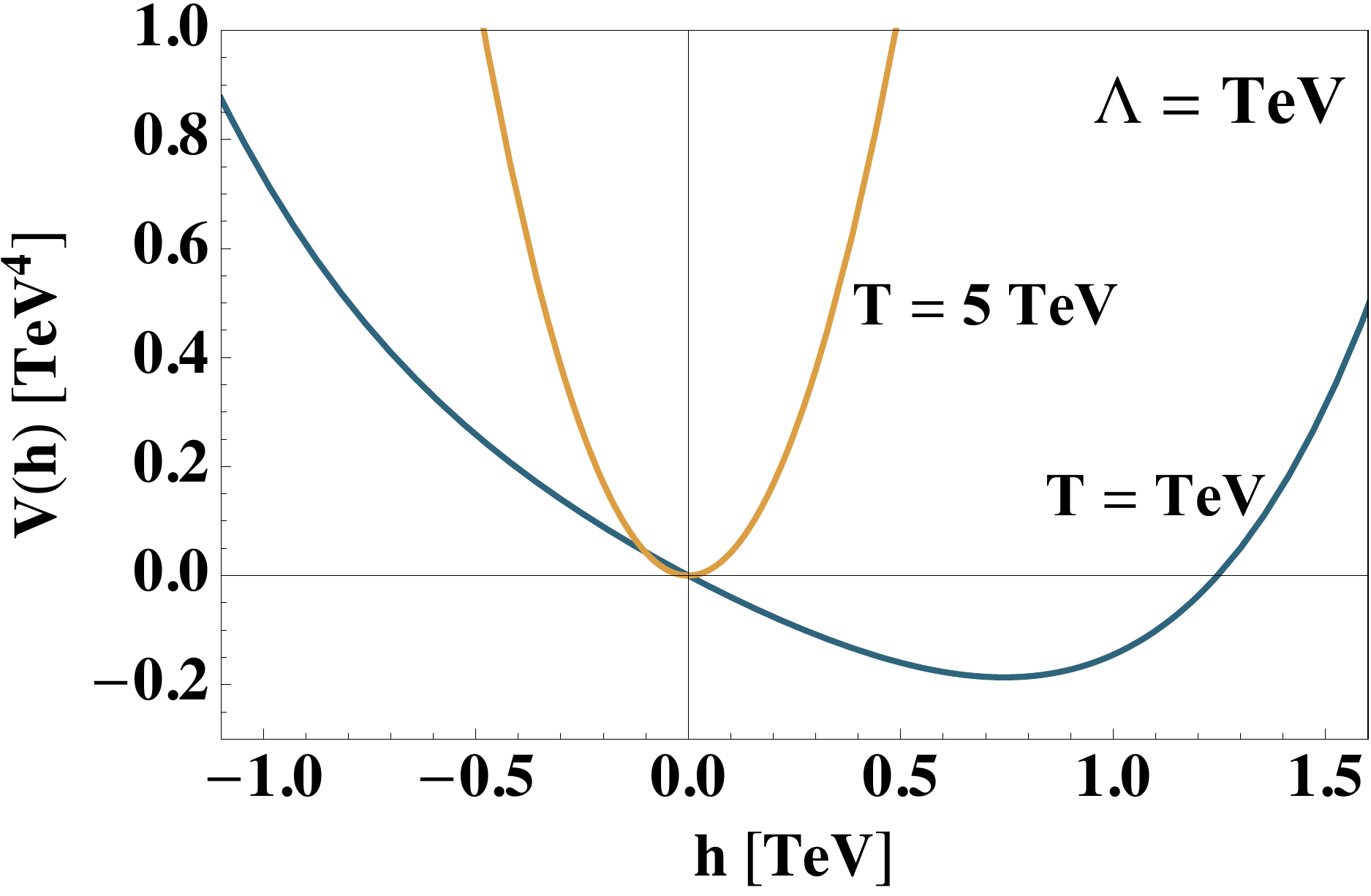}
\caption{Finite-temperature Higgs potential at $T= 1,5$~TeV,
where the chiral effective theory cut-off is taken to be $\bar{\Lambda}=4$~TeV.} \label{fig:finiteThiggs}
\end{figure}

The most important entry in $Y$ is the 66 entry corresponding to the top quark with $y_{66} = y_t \sim 1$.  
Including the tadpole generated by $\langle \Pi_{66}\rangle$ in the finite temperature Higgs potential gives
\begin{align}
 V_{\rm eff}(h, T)\simeq& -\frac{\bar{\Lambda}^2}{4\pi}y_t \langle \Pi_{66}\rangle \, h 
+\left(\frac{\alpha}{24}(T^2-T_c^2)\right)h^2\notag \\
& -\gamma T h^3 +\lambda h^4, \label{eq:Vheff} 
\end{align}
where
\begin{align*}
&\alpha = \sum_{\rm bosons}g_i\left(\frac{m_i}{v}\right)^2 + \frac12 \sum_{\rm fermions}g_i\left(\frac{m_i}{v}\right)^2, \\
&T_c=2v\sqrt{\frac{6\lambda}{\alpha}},~~~
\gamma=\frac{\sqrt{2}}{6\pi}\sum_{\rm bosons}g_i\left(\frac{m_i}{v}\right)^3,
\end{align*}
and the sums go through all massive bosons and fermions in the SM. 
Taking $\langle \Pi_{66}\rangle \sim {\bar{\Lambda}}/{4\pi}$, the Higgs vev can be well approximated by
\begin{align}
v \simeq\frac{\bar{\Lambda}}{4}\left( \frac{y_t}{\pi^2\lambda} \right)^{1/3}. 
\end{align}
For $\Lambda \sim 1$~TeV and taking $\bar{\Lambda}=4~$TeV, the resulting Higgs VEV 
is $v\simeq 0.9~$TeV, as can be seen in Fig.~\ref{fig:finiteThiggs}.

\section{Applications}

An early period of QCD confinement can have profound implications 
for the history of the early universe.  We flesh out one particularly exciting possibility to generate the observed baryon asymmetry
of the Universe, and sketch several more which would be worth following up in future work below.

\subsection{Baryogenesis}

QCD confining at $\sim$~TeV temperatures combined with the axion as a solution to the strong CP problem allows for a novel mechanism
to explain the baryon asymmetry of the Universe.  As mentioned above, since confinement at a TeV scale takes place when all six of the
SM quarks are massless, the phase transition is expected to be first order \cite{PhysRevD.29.338, 1742-6596-432-1-012027} and proceeds through bubble nucleation. 
Inside the bubble QCD is confined and the EW symmetry is broken (and thus baryon-number violation through the weak interaction
is inoperative) whereas outside remains in the unbroken and unconfined phases. 
Furthermore, if there exists an axion field that addresses the strong CP problem, there can be 
large CP violation from the uncancelled strong phase 
during this phase transition \cite{Kuzmin:1992up, Servant:2014bla}.

The axion couples to the baryon current through the interactions of the pseudoscalar $\eta'$ meson,
whose mass scales like $m_{\eta'} \sim \Lambda$. 
At energies lower than the $\eta'$ mass, its residual effects are described by the effective Lagrangian:
\begin{align}
\mathcal{L}_{\rm eff} \simeq \frac{10}{f_{\pi}^2 m_{\eta'}^2}\frac{\alpha_s}{8\pi} G\widetilde{G} \frac{\alpha_w}{8\pi}  W\widetilde{W},
\end{align}
where $W$ ($\widetilde{W}$) is the $SU(2)_W$ (dual) field strength.
As the axion rolls to its minimum, there is an uncancelled $\bar{\theta}$ 
which induces a $G\tilde{G}$ condensate \cite{Kuzmin:1992up},
\begin{align}
\frac{\alpha_s}{8\pi} \langle G\tilde{G}\rangle = m_a^2(T) f_a^2 \sin\bar{\theta}(T).
\end{align}
Through the anomaly equation, $\partial_\mu j_B^\mu = \alpha_W / 8 \pi W \widetilde{W}$, $W \widetilde{W}$
is related to the baryon current density $j_B^\mu$.  Integrating by parts produces an effective 
chemical potential $\mu$ for baryons:
\begin{align}
\mu=\frac{d}{dt}\left[ \frac{10}{f_\pi^2 m_{e\eta'}^2}m_a^2(T)f_a^2 \sin\bar{\theta}(T)\right].
\end{align}

The non-zero chemical potential drives production of a non-zero baryon asymmetry by the electroweak sphalerons, 
\begin{align}
n_B =\int_{T_i}^{T_f} dt  \frac{\Gamma_{\rm sph}(T)}{T} \mu, 
\end{align}
where $\Gamma_{\rm sph}(T) \sim 25\, \alpha_w^5\,T^4$ is the thermal sphaleron rate outside the bubble where $\langle h\rangle=0$. 
We assume that inside the bubble the sphalerons are sharply switched off.
Making use of the temperature-dependent axion mass 
\cite{PhysRevD.33.889}
\begin{align}
&m_a^2(T)\simeq m_a^2(T=0) \times \left(\frac{\Lambda}{T}\right)^7, \\
&{\rm where}~~~m_a^2(T=0) \sim m_\pi^2 ~\frac{f_\pi^2}{f_a^2}, \notag
\end{align}
assuming $\sin\bar{\theta}$ is varying slowly, 
and $\Delta m_a^2(T_{\rm PT})\simeq m_a^2(T_{\rm PT})$ around $T_{\rm PT}$, the temperature at which the EW transition happens, the resulting baryon-to-entropy ratio is
\begin{align}
\eta =\frac{n_B}{s}\simeq \frac{45\times 125}{2\pi^2 g_*(T_{\rm reh})}\frac{m_\pi^2}{m_{\eta'}^2} \alpha_w^5\sin\bar{\theta} \left(\frac{T_{\rm PT}}{T_{\rm reh}}\right)^3 \left(\frac{\Lambda}{T_{\rm PT}}\right)^7,
\end{align}
where $T_{\rm reh}$ is the reheat temperature at the end of the EW phase transition and $g_*$ counts 
the relativistic degrees of freedom at that time. This picture has all of the dynamics naturally occurring concurrently, resulting in 
$T_{\rm reh}\simeq T_{\rm PT}\simeq \Lambda$ and in the baryon asymmetry being
roughly independent of the temperature at which QCD confines provided this happens above the electroweak scale. 
The baryon-to-entropy ratio is
\begin{align}
\eta \simeq 1.8\times 10^{-9}\times \sin\bar{\theta},
\end{align}
to be compared with the Planck measurement \cite{Ade:2015xua}, 
\begin{align}
\eta_{\rm exp}=(8.59\pm 0.11)\times 10^{-11}.
\end{align}
Achieving the observed baryon asymmetry requires a very modest tuning of $\sin \bar{\theta} \sim 1/10$
or a small amount of dilution after the baryon asymmetry is generated.  It is remarkable that the baryon asymmetry is naturally close to the observed value for $\bar{\theta}$ of order one, despite the relative dearth of adjustable parameters.

\subsection{Other Applications}

The baryon asymmetry is just one application of an early period of QCD confinement out of many that could be imagined.  We leave detailed follow up for future work, but a few others would be:

\emph{BBN and Early universe.} Early QCD confinement could leave an imprint on BBN if the transition to $\Lambda \sim 1$~GeV occurs late enough, implying bounds on the dynamics of $\phi$. Furthermore, while confined the SM plasma degrees of freedom are different, influencing the evolution of the Universe.

\emph{Dark Matter Freeze-out.} If dark matter freezes out during a period in which QCD is confined, the relevant degrees of freedom both for annihilation and in the plasma correspond to the confined phase, in contrast with usual WIMP scenarios.

\emph{Axion Dark Matter.} For theories invoking a QCD axion, the early period of confinement switches on the axion 
potential earlier and generates a larger axion mass (of order $10^4$ the usual mass for $\Lambda\sim$~TeV). 
For very large $\Lambda$, the axion could decay on cosmological time scales, erasing its density.  Even for more modest $\Lambda$, the transition to
 $\Lambda \sim 1$~GeV would induce a novel temperature-dependence on the axion mass, and could e.g. result in an early period of matter domination.

\emph{Gravitational Waves.} As with other first order cosmological phase transitions, the early period of QCD confinement is expected to generate gravitational waves \cite{Caprini:2015zlo, Cutting:2018tjt}. The detailed predictions will depend sensitively on the dynamics of the bubble nucleation, expansion, and collision,
which themselves take place in a background of the strongly interacting plasma (which could, for example, provide friction slowing down the bubble expansion rate).
A careful investigation of the properties of the phase transition and its impact on gravitational wave production is
currently under investigation \cite{GRwaves}.

\emph{Collider searches.}  The most model-independent prediction is the existence of the neutral scalar field $\phi$ which couples to gluons, and could contribute
to dijet signatures at high energy colliders.  If it mixes with the Higgs, it will pick up other couplings to SM fields, and induce deviations in the Higgs couplings. 
There may be additional neutral or colored particles coupled to $\phi$ as well.

\emph{Heavy Ion Collisions.} If the dynamics restoring $\Lambda \sim 1$~GeV occur at low energies, there are likely to be indications visible in high temperature
environments such as heavy ion collisions.

\emph{$SU(2)_W$ Confinement.} A similar mechanism could be employed, e.g. to trigger $SU(2)_W$ confinement in the early Universe, with interesting consequences for electroweak symmetry breaking and potentially opening more new avenues for baryogenesis.

\section{Conclusions and Outlook}
\label{sec:conclusions}

Given our lack of knowledge about QCD at high temperatures, it is natural to ask whether there may be surprising dynamics which
were important in the early Universe, but remain hidden at low temperatures.  If QCD confines at a high scale,
returning to $\Lambda \sim 1$~GeV at later times, it may shed light on some of the
mysteries of our Universe, including the fact that it is made out of matter and not anti-matter.
We have sketched the basic properties of such a scenario, and demonstrated that baryogenesis can work if there is an axion
which solves the strong CP problem.  
Many open questions remain open, and many avenues remain to be explored in this framework.

\section*{Acknowledgements}

We are grateful for discussions with Djuna Croon, Anna Hasenfratz, Christopher T. Hill, Ciaran Hughes, 
David B. Kaplan, David McKeen, Michael Ramsey-Musolf, and Geraldine Servant. This work is supported in part by NSF Grant No.~PHY-1620638.
SI acknowledges support from the UC Office of the President via a 
UC Presidential Postdoctoral fellowship. This work was performed in part at Aspen Center for Physics, 
which is supported by National Science Foundation grant PHY-1607611. 

\bibliography{ref}

\end{document}